# A Hyperbolic Decay of the *Dst* Index during the Recovery Phase of Intense Geomagnetic Storms

J. Aguado, C. Cid, E. Saiz, and Y. Cerrato

Space Research Group- Space Weather Science, Departamento de Física,

Universidad de Alcalá, Alcalá de Henares, Spain

**Abstract.** What one commonly considers for reproducing the recovery phase of magnetosphere, as seen by the *Dst* index, is exponential function. However, the magnetosphere recovers faster in the first hours than in the late recovery phase. The early steepness followed by the late smoothness in the magnetospheric response is a feature that leads to the proposal of a hyperbolic decay function to reproduce the recovery phase, instead of the exponential function. A superposed epoch analysis of recovery phases of intense storms from 1963-2006 was performed, categorizing the storms by their intensity into five subsets. The hyperbolic decay function reproduces experimental data better than what the exponential function does for any subset of storms, which indicates a non-linear coupling between *dDst/dt* and *Dst*. Moreover, this kind of mathematical function, where the degree of reduction of the *Dst* index depends on time, allows for explaining different lifetimes of the physical mechanisms involved in the recovery phase and provides new insights for the modeling of the *Dst* index.

## 1. Introduction

As a result of the solar wind-magnetosphere coupling, there is energy transfer into the inner magnetosphere. Plasma sheet ions were thought for many years to be energized and trapped on closed drift paths producing a symmetric ring current around the Earth. The strength of the ground disturbance produced by the gyration and drift of these ions was quantified by the hourly *Dst* index [Sugiura and Kamei, 1991], calculated by averaging horizontal magnetic deviations observed at four low latitudes stations. This index was considered as a measure of the ring current intensity reporting on the total energy of ring current particles through the Dessler-Parker-Sckopke (DSP) relation [Dessler and Parker, 1959; Sckopke, 1966].

Looking at *Dst* index, the main feature of a geomagnetic storm is a depression, corresponding to the main phase of the storm, lasting between approximately 3 and 12 hours, which is followed by a slower recovery during which *Dst* increases back toward zero over hours to tens of hours (recovery phase) because of the ring current decay. The minimum value reached by *Dst* index corresponds to the peak value and it is considered as a magnitude of the intensity of the storm, so that a storm is considered intense if *Dst* peak value reaches at least -100 nT [Gonzalez et al., 1994].

Nowadays it is also considered that the ring current is the dominant contributor to the *Dst* index, although it is influenced from other current systems such the magnetopause, magnetotail and induced Earth currents. However, the idea of a symmetric ring current remains only for the late recovery phase. As energetic ions from the plasma sheet are convected deep into the dipolar regions under the action of enhanced convection electric field, an intense asymmetric ring current (partial ring current) develops. The injection model, first proposed by DeForest and McIlwain [1971], predicted that the ring current was asymmetric only as long as injection continues, that is in the main phase of the storm. However, it is now understood that the partial ring current far exceeds the symmetric ring current throughout the entire main phase and into the very early recovery phase of moderate and intense geomagnetic storms. Several papers have considered this issue from a theoretical point of view [e.g., Takahashi et al., 1990; Ebihara and Ejiri, 1998, 2000; Jordanova et al., 1998; Liemohn et al., 1999, 2001; Kozyra et al., 2002; Kozyra and Liemohn, 2003; Liemohn and Kozyra, 2005] and from an observational one [e.g., Greenspan and Hamilton, 2000; Jorgensen et al., 2001; Mitchell et al., 2001; Pollock, 2001; Soraas et al., 2002, 2003]. The asymmetric ring current is consequence of the energetic injected ions which move on open drift paths once through the inner magnetosphere before they pass through dayside magnetopause

[Liemohn et al., 1999, 2001; Kozyra et al., 2002; Daglis and Kozyra, 2002; Fok et al., 2003; Burch, 2005; Kalegaev et al., 2008]. As the early recovery phase of the storm begins, the convection electric field weakens. This decrease turns open drift paths into closed ones forming the symmetric ring current. At the end of the early recovery phase, ~80-90% of the remaining ring current energy is trapped in closed drift paths [Daglis and Kozyra, 2002]; a major symmetric ring current component only appears in the late recovery phase [Liemohn and Kozyra, 2005].

Loss of the storm-time ring current energy (and thus recovery of the $Dst$ index toward zero) was believed to occur dominantly through charge exchange with the neutral hydrogen geocorona. In fact, the decay of large magnetic storms was split in two-phases: a early fast recovery followed by a slower one, which was believed to be the result of the large differences between the charge exchange lifetimes of oxygen and hydrogen ions with energies above 50 keV [Tinsley and Akasofu, 1982; Hamilton et al., 1988]. The much more rapid removal of oxygen ions was thought to be the cause of the fast loss lifetimes during the early recovery phase. By the end of the early recovery phase, the ring current was significantly depleted in oxygen ions relative to protons. The long charge-exchange lifetimes of the proton component dominated the late recovery phase. The preferential removal of oxygen ions by charge exchange in the early recovery phase was thought to drive the observed dramatic composition changes that were correlated closely with the ring current recovery [Hamilton et al., 1988; Daglis, 1997]. Daglis et al. [2003] argued that differential charge exchange loss between hot oxygen ions and hot hydrogen ions (rapid for the first one and slower for the second) was a major factor in the two-phase decay recovery for some storms.

Trying to explain the significant recovery of the $Dst$ index in the early recovery phase, Feldstein et al. [2000] and Ohtani et al. [2001] argued that it could be related to a rapid shut off the tail current. However, O'Brien et al. [2002] statistically analyzed the recovery rate of *Dst* for storms with rapid shut-off of the convection strength versus those with gradual shut-off (continued convection) and they found that the two groups of storms had statistically identical decay rates.

The changeover from rapid removal at the dayside magnetopause during the main and early recovery phases to much slower charge exchange removal of trapped ring current particles during the late recovery phase were also proposed to account for the two distinctly different lifetimes that dominate the ring current recovery [Jordanova et al., 2003; Kozyra and Liemohn, 2003]. That is, continued convection into the recovery phase caused the initial fast recovery of the ring current, and a rapid shut-off of this flow-out suddenly stopped this loss process, allowing the slower loss processes to dominate the recovery time scale.

Other loss processes were also proposed as contributors to the storm-time ring current decay: Coulomb collisions between the hot ring current ions and plasmaspheric particles [Fok et al, 1991, 1993, 1995, 1996; Jordanova et al., 1998] and ion precipitation into the upper atmosphere due to the strong pitch angle scattering of particles into the loss cone by wave-particle interactions (especially EMIC waves) [Kozyra et al., 1997; Jordanova et al., 1997, 2001]. Walt and Voss [2001] concluded that wave-particle interactions elevate particle precipitation losses to a level capable of producing a rapid initial recovery of the ring current. However, Kozyra et al. [1998, 2002] and Liemohn et al. [1999] stated that although the removal of ions from open drift paths by charge exchange interactions and precipitation decreased the ring current lifetime even further, these were secondary effects. Other studies have shown that wave-induced particle precipitation is a minor component of the total loss rate from the ring current [e.g., Jordanova et al., 1998, 2001; Soraas et al., 2002, 2003; Khazanov et al., 2002, 2003].

Liemohn and Kozyra [2005], based on idealized simulations of ring current decay, concluded that differential charge exchange loss rate of hot $O^+$ and hot $H^+$ could not produce a two-phase decay. However, they showed that a two-phase decay can only be created by the transition from flow-out to charge exchange dominance of the ring current loss. They also showed that flow-out loss was the only process with sufficient intensity and variability to cause a sudden increase in the ring current energy loss lifetime.

On the other hand, a number of studies have previously examined the decay time of both single and double exponential fits to the recovery phase of Dst index [Burton et al., 1975; Hamilton et al., 1988; Ebihara et al., 1998; Dasso et al., 2002; Kozyra et al., 2002; Weygand and McPherron, 2006; Monreal MacMahon and LLop, 2008]. The exponential fits are based on the assumption of decay rate of the ring current is proportional to the energy content of the ring current (through the DPS relation), that is, on a linear dependence of the *dDst/dt* upon *Dst*. In doing so, the temporal evolution of *Dst* index (after correcting from magnetopause and magnetotail currents) is modeled in terms of an injection function, $Q(t)$, and a recovery characteristic time scale, $\tau$, leading to an exponential decay for the corrected *Dst* index.

Different recovery characteristic times have been proposed. Burton et al. [1975] proposed a constant value of 7.7 hours. Fenrich and Luhmann [1998] first considered the influence of the convective electric field ($E_y$) on the recovery time and proposed two different $\tau$ values (3 and 7.7 hours) depending on whether the magnitude of $E_y$ was lower or greater than 4 mV/m, respectively. Revising this relationship, O´Brien and McPherron [2000] proposed an expression for $\tau$ as a function of $E_y$. Maintaining as an aim the accuracy in reproducing the recovery phase of *Dst* index, Wang et al. [2003]

proposed that the $\tau$ dependence on solar wind was not only related to $E_y$ but also to dynamic pressure.

The influence of the intensity of the storm on the recovery time has also been studied. Prigancová and Feldstein [1992] distinguished two stages in the recovery phase with two different $\tau$ values: $\tau = 1$ hour ($\tau = 0.5$ hours for the most intense storms) for the early stage of the storm recovery phase and $\tau = 5 - 10$ hours for the late stage. More recently, Dasso et al. [2002] proposed a mean value of $\tau = 14 \pm 4$ hours, which decreased with the intensity of the storm. However, there was no empirical or theoretical function that quantified the dependence of $\tau$ with the intensity of the storm.

Against this backdrop, a new proposal is made in this paper to model the recovery phase of geomagnetic storms, as seen by *Dst* index, based on a new decay function that better fits experimental data and considers the dependence of the recovery time on the intensity and time. This new function, the hyperbolic decay, is consistent with the loss processes associated with different lifetimes at different stages and different storm intensities, as described above.

## 2. Recovery phase modeling: exponential function versus hyperbolic decay function

Exponential decay function, $Dst(t) = Dst_0 e^{-t/\tau}$, commonly used to model the recovery phase of geomagnetic storms, assumes that the degree of reduction of *Dst*, defined as $-(dDst/dt)/Dst$, is independent of time and of $Dst_0$ (minimum value of *Dst* index). In fact, the degree of reduction of exponential function is $1/\tau$, $\tau$ being the characteristic recovery time. However, as described above, different decay processes are involved at different stages of the recovery phase of a magnetic storm, and therefore in *Dst* index.

On the other hand, different recovery times have been proposed in literature depending on the intensity of the storm. Therefore, a recovery characteristic time, dependent on time and $Dst_0$, would be expected.

A notable distinction exists between exponential function and hyperbolic decay function in so far as the degree of reduction of the decaying magnitude (in this case *Dst* index) is concerned. If *Dst* in the recovery phase of a geomagnetic storm is described by the hyperbolic decay function as $Dst(t) = \frac{Dst_0}{1 + t/\tau_h}$, the degree of reduction of *Dst*, as defined above, is $\frac{1}{\tau_h + t}$. Thus, the degree of reduction of the hyperbolic decay function decreases monotonously with time, instead of being a constant value ($1/\tau$) as in the exponential decay one.

Other key difference arises considering the modeling of temporal evolution of *Dst* index by a hyperbolic law instead of an exponential one: if the coupling of the *dDst/dt* upon *Dst* is linear, then, it results in an exponential decay law, but if the coupling becomes non-linear, that is $dDst/dt \propto Dst^2$, then, hyperbolic law represents the corresponding solution of the problem [Pop, F. A., Li, K. H., 1993].

Concerning the meaning of the parameters involved in both decay functions: hyperbolic and exponential, it is important to note that both, approach zero value when time goes to infinite and the same value ($Dst_0$) when time goes to zero, that is, to the intensity of the storm. As a result, the meaning of the parameter, $Dst_0$, is the same for the two decay functions: the initial value of the function. On the other hand, the meaning of the corresponding 'recovery time' ($\tau$ or $\tau_h$) differs from the exponential function to the hyperbolic function. In the first one, $\tau$ represents the time needed to reach initial value/*e*, while for the second one $\tau_h$ represents the time needed to reach initial value/2. Thus, although both recovery times have a different meaning, for comparison purposes in the context of the previous studies, it would be useful to consider the time needed for hyperbolic function to reach initial value/$e$, that is, $\tau_h(e-1)$.

An outstanding difference between hyperbolic and exponential decay arises when both functions are supposed to reproduce experimental data which reach 1% of the initial value (which is comparable to the end of the decay) for a fixed time interval. In doing so, the exponential function will last a time $t = 4.6\tau$ while the hyperbolic function will need $t = 99\tau_h$. As the time interval is fixed, it should be the same for both functions, and then, the relationship between both recovery times is $\tau_h \approx 0.05\ \tau$. As a consequence, the curvature of the hyperbolic function (obtained as the inverse of the second derivative of the function) at initial stages is $1.25\times10^{-3}$ times less than the curvature of the exponential function, which evidences that the hyperbolic function will provide a steeper response than the exponential function for decaying 99% of the initial value during the same time interval.

## 3. Selection of Storms and Superposed Epoch Results

Every intense storm ($Dst < -100$ nT), from 27th November 1963 to 31th December 2003, is considered for this study. This period includes all definitive $Dst$ data available from the World Data Center of Geomagnetism, Kyoto, at http://swdcwww.kugi.kyoto-u.ac.jp/.

Recovery phases, starting at $Dst_{peak}$ (minimum of $Dst$), has been analyzed to select the 'pure recovery' events. Therefore, those storms with dips that arise during the recovery phase, which indicates that a substantial injection of energy is taking place, have been excluded from the analysis. However, storms with several dips in the main phase of the storm, that is, before the $Dst_{peak}$ value is achieved, have been considered for this study.

In so far as the significance of the dip is concerned, the relative amount of energy input during the recovery phase between different *Dst* dips is considered the most appropriate signature to check if the event can be considered a 'pure recovery' event. The criterion applied is that a negligible injection of energy is taking place when the dip does not exceed 15 per cent of $Dst_{peak}$ value.

Finally, 148 storms from 1967-2003, which do not include substantial injection of energy during the recovery phase, are included in this study.

A superposed epoch analysis of recovery phases of geomagnetic storms has been conducted using zero as the epoch time for the $Dst_{peak}$ of every storm, and by extending the epoch time to 48 hours. To analyze, not only the temporal dependence of the recovery time, but also the intensity dependence, several subsets have been made of the set of 148 storms based on their intensity. Four subsets, defined by the $Dst_{peak}$, have been made with a dynamic range from -100 nT to -300 nT, decrementing each subset by 50 nT, that is (-100 nT, -150 nT], (-150 nT, -200 nT], (-200 nT, -250 nT], (-250 nT, -300 nT]. The subset number 5 includes all the storms whose $Dst_{peak}$ values are lower than -300 nT. The storms of each subset have been averaged and the mean recovery phase obtained. Figure 1, which shows the averaged time histories of recovery phases for different subsets, evidences that the recovery phase depends on the intensity of the storm.

Exponential (blue dashed line) and hyperbolic decay (red solid line) fittings have been plotted along with mean recovery phase for the five storm subsets (Figure 2). The exponential fittings of the five mean recovery phases show similar features. Although all of them seem to fit well, considering the $r^2$ value (always bigger than 0.92), the exponential curve is always above the experimental data during the first 4-6 hours

(epoch time) and from 30 hours of the recovery phase; otherwise, it is under experimental data. This indicates that the recovery of magnetosphere is faster than that of the exponential function during the initial stage, and slower during the late stage, suggesting thereby a hyperbolic decay function to explain the evolution of *Dst*.

Figure 2 also proves that a hyperbolic function is a better approach than an exponential function for experimental data. From the values of $r^2$ over 0.99 for every mean storm, one can conclude that the magnetosphere recovers as a hyperbolic function, with a degree of reduction of *Dst* that decreases in time.

Figure 3 shows a scatter plot of the parameters obtained from the fitting of the hyperbolic function for each mean recovery phase of different subsets: $\tau_h$ versus $Dst_0$. At a first glance, Figure 3 suggests a linear dependence between the recovery time, $\tau_h$, and the intensity of the storm. A linear fitting provides the regression function $\tau_h = (20 \pm 1) + (0.029 \pm 0.005) Dst_0$, with $r^2=0.92$. The lowering of the $r^2$ in the curve is related to the deviation of the point (-218.3 nT, 12.37 hours), corresponding to the subset including those geomagnetic storms with $Dst_{peak}$ value between -200 nT and -250 nT (subset #3). This fact is evidenced by the new linear fitting removing this point from the regression (dashed line in the Figure 3), where the $r^2$ value increases until 0.999. We have revised the 13 events included in the subset #3 modifying the criterion for a negligible injection of energy to dips which do not exceed 5 per cent of $Dst_{peak}$ value. Only three events remain in the new subset. The new $\tau_h$ value obtained from the superposed epoch analysis of these three events has been plotted in the Figure 3 with a plus symbol. As can be seen, the new point follows the trend of the other points included in the graph and is close to the dashed line, corresponding to the linear regression with higher $r^2$ value.

Although might be tempting a revision of the whole analysis made in this paper, modifying the criterion for a negligible injection of energy to dips which do not exceed 5 per cent of $Dst_{peak}$ value will not be statistically reliable because of the drastic reduction in the number of events (from 148 to 26, including the five subsets). An increase in the number of events available throughout the next years will allow to revise this work including a larger sample.

## 4. Summary and Conclusions

The authors have studied all the intense ($Dst \leq$ -100 nT) storms from 1963 to 2003 that exhibited a negligible injection of energy during their recovery phase. Based on a superposed epoch analysis, the study demonstrates that the recovery of the magnetosphere is hyperbolic, rather than exponential. From Figure 2 we show that the hyperbolic decay reproduces accurately experimental data in every subset, although the recovery time changes from one subset to another. Moreover, the hyperbolic recovery times are linearly related to the initial values of the $Dst$ index for every subset (see Figure 3). Therefore, we can conclude that the recovery of the bulk magnetosphere after an intense energy transfer from solar wind follows a hyperbolic law, with a degree of reduction of $Dst$ depending on time, and where the recovery time depends linearly on the intensity of the storm.

The recovery time values, obtained for the averaged storms of different subsets, range between 10.4 to 16.8 hours (see Figure 3), decreasing linearly with the intensity of the storm. Although these recovery time values are similar to those proposed in literature for the exponential function decay time [e.g. Burton et al., 1975; Dasso et al., 2002; Wang et al., 2003; O'Brien and McPherron, 2000], both recovery times are not comparable magnitudes.

The above results, which demonstrate that the hyperbolic decay function fits accurately the recovery of the magnetosphere, should be used to address the physical mechanisms involved in the recovery phase of geomagnetic storms. This problem has been dealt previously with a two-phase decay (or even more), trying to fit the different stages by different exponential functions, as stated above. The hyperbolic function is able to embrace the appearance and disappearance of different physical processes in a gradual way and with only one function for the complete recovery phase. In this way, the dependence on time of the degree of reduction of *Dst* magnitude makes the hyperbolic function able to explain the existence of diverse non-linearly coupled loss processes during the recovery of the magnetosphere.

As a consequence, it is possible to explain that at the early recovery phase the main mechanism involved is the flow-out loss (although the other loss processes are also involved), being charge exchange the only mechanism that survives at the late stage. Moreover, differential charge exchange loss rate of hot $O^+$ and hot $H^+$ ions changing with epoch time can also be included in a hyperbolic decay function, even if the different contributions cannot be separated. As pointed out by Liemohn and Kozyra [2005], charge exchange loss lifetimes depend on the ion energy and the radial distance *L*. In this way, although at high energies $O^+$ ions have shorter lifetimes than protons, and then there would be expected a large loss rate of $O^+$ contributing significantly only early in the recovery phase, at low energy range injected $H^+$ ions, will be rapidly exchanged, making $H^+$ loss rate comparable to that of $O^+$. As significant levels of low-energy $H^+$ ions are present throughout the recovery phase and the ring current extends to a wide range of L values, a sudden change is not expected.

The accuracy of the hyperbolic fitting in reproducing the recovery phase of *Dst* index addresses, not only the existence of diverse processes involved in a gradual way, but also the diverse nature of the processes involved: flow-out, charge exchange, particle precipitation by wave-particle interaction, etc. This diverse nature suggests a non constant degree of reduction of *Dst* index and then, a non-linear coupling of *dDst/dt* upon *Dst*.

One of the important outcomes of our study is the proposal of a unique continuous function to model the magnetospheric response after a huge injection of energy from solar wind, which is a great improvement in the modeling of the *Dst* index as a function of time. This hyperbolic decay function denotes a steeper response in the early recovery phase which let reproduce the observations for intense and severe storms (the aim of this paper) widely related in literature.

Concerning the relationship between the recovery time and the intensity of the storm (or *Dst* peak value), it was proposed its existence and different values for the recovery time were proposed for different intensity intervals [e.g. Monreal MacMahon and LLop, 2008 and references therein]. It was also reported [e.g. Mendes Jr., 1992] that the decay time, considering *Dst* intervals, results on discontinuities in the relation between the ring current dissipation and the coupling function. Instead of a discontinuous function, our results, as shown in Figure 3, provide a continuous function of *Dst* peak value to compute the recovery time.

In summary, this paper provides a new continuous function to reproduce the entire recovery phase of the magnetosphere, as seen by *Dst* index. The fact that a hyperbolic law represents the corresponding solution of the temporal evolution of *Dst* index means that the coupling of *dDst/dt* upon *Dst* is non-linear. Although the physical implications of this dependence are still in their beginning, we sense that in the light of these results a new generation of models will rise for the temporal evolution of the *Dst* index based on the energy balance in the ring current. The replacement of the loss term proportional to the own *Dst* index by a non-linear term related to the hyperbolic decay function proposed above is out from the scope of this paper, but will be our aim in a future work.

## Acknowledgments

This work was supported by grants ESP 2006-08459 and AYA2009-08662 from the Ministerio de Ciencia e Innovación of Spain.

# References

Burch, J. L. (2005), Magnetospheric imaging: Promise to reality, *Rev. Geophys.*, *43*, RG3001, doi:10.1029/2004RG000160.

Burton, R. K., R. L. McPherron, and C. T. Russell (1975), An Empirical Relationship between Interplanetary Conditions and Dst, *J. Geophys. Res.*, *80*(31), 4204–4214.

Daglis, I. A. (1997), The Role of Magnetosphere-Ionosphere Coupling in Magnetic Storm Dynamics, in Magnetic Storms, *Geophys. Monogr. Ser.*, 98, edited by B. T. Tsurutani et al., p. 107, AGU, Washington D. C.

Daglis, I. A., and J. U. Kozyra (2002), Outstanding issues of ring current dynamics, *J. Atmos. Terr. Phys., 64*(2), 253-264, doi: 10.1016/S1364-6826(01)00087-6.

Daglis, I. A., Kozyra, J. U., Kamide, Y., Vassiliadis, D., Sharma, A. S., Liemohn, M. W., Gonzalez, W. D., and Tsurutani, B. T. (2003), Intense space storms: Critical issues and open disputes, J. *Geophys. Res.*, *108*(A5), 1208, doi:10.1029/2002JA009722.

Dasso, S., D. Gómez, and C. H. Mandrini (2002), Ring Current Decay Rates of Magnetic Storms: A Statistical Study from 1957 to 1998, *J. Geophys. Res.*, *107*(A5), 1059, doi:10.1029/2000JA000430.

DeForest, S., and C. E. McIlwain (1971), Plasma clouds in the magnetosphere, *J. Geophys. Res.*, *76*(16), 3587–3611.

Dessler, A. J., and E. N. Parker (1959), Hydromagnetic Theory of Geomagnetic Storms, *J. Geophys. Res.*, *64*(12), 2239–2252.

Ebihara, Y., and M. Ejiri (1998), Modeling of solar wind control of the ring current buildup: A case study of the magnetic storms in April 1997, *Geophys. Res. Lett.*, *25*(20), 3751-3754.

Ebihara, Y., M. Ejiri, and H. Miyaoka (1998), Coulomb lifetime of the ring current ions with time varying plasmasphere, *Earth Planets Space*, *50*(4), 371-382.

Ebihara, Y., and M. Ejiri (2000), Simulation study on fundamental properties of the storm-time ring current, *J. Geophys. Res.*, *105*(A7), 15843-15859.

Feldstein, Y. I., L. A. Dremukhina, U. Mall, and J. Woch (2000), On the two-phase decay of the Dst-variation, *Geophys. Res. Lett.*, *27*(17), 2813-2816, doi: 10.1029/2000GL003783.

Fenrich, F. R., and J. G. Luhmann (1998), Geomagnetic Response to Magnetic Clouds of Different Polarity, *Geophys. Res. Lett.*, *25*(15), 2999–3002.

Fok, M.-C., J. U. Kozyra, A. F. Nagy, and T. E. Cravens (1991), Lifetime of Ring Current Particles due to Coulomb Collisions in the Plasmasphere, *J. Geophys. Res.*, *96*(A5), 7861–7867.

Fok, M.-C., J. U. Kozyra, A. F. Nagy, C. E. Rasmussen, and G. V. Khazanov (1993), Decay of equatorial ring current ions and associated aeronomical consequences, *J. Geophys. Res.*, *98*(A11), 19381-19393.

Fok, M.-C., T. Moore, J. Kozyra, G. Ho, and D. Hamilton (1995), Three-Dimensional Ring Current Decay Model, *J. Geophys. Res.*, *100*(A6), 9619-9632.

Fok, M.-C., T. Moore, and M. Greenspan (1996), Ring current development during storm main phase, *J. Geophys. Res.*, *101*(A7), 15311-15322.

Fok, M.-C., et al. (2003), Global ENA image simulations, *Space Sci. Rev., 109*, 77–104.

Gonzalez, W.D., J.A. Joselyn, Y. Kamide, H.W. Kroehl, G. Rostoker, B. T. Tsurutani, and V. M. Vasyliunas (1994). What is a geomagnetic storm?, *J. Geophys. Res.*, *99*(A4), 5571-5792.

Greenspan, M. E., and D. C. Hamilton (2000), A test of the Dessler-Parker-Sckopke relation during magnetic storms, *J. Geophys. Res., 105*(A3), 5419–5430.

Hamilton, D. C., G. Gloeckler, F. M. Ipavich, W. Stüdemann, B. Wilken, and G. Kremser (1988), Ring Current Development during the Great Geomagnetic Storm of February 1986, *J. Geophys. Res.*, *93*(A12), 14343–14355.

Jordanova, V., J. Kozyra, A. Nagy, and G. Khazanov (1997), Kinetic model of the ring current-atmosphere interactions, *J. Geophys. Res.*, *102*(A7), 14279-14291.

Jordanova, V., C. Farrugia, L. Janoo, J. Quinn, R. Torbert, K. Ogilvie, R. Lepping, J. Steinberg, D. McComas, and R. Belian (1998), October 1995 magnetic cloud and accompanying storm activity: Ring current evolution, *J. Geophys. Res.*, *103*(A1), 79-92.

Jordanova, V. K., C. J. Farrugia, R. M. Thorne, G. V. Khazanov, G. D. Reeves, and M. F. Thomsen (2001), Modeling Ring Current Proton Precipitation by Electromagnetic Ion Cyclotron Waves during the May 14–16, 1997 Storm, *J. Geophys. Res.*, *106*(A1), 7–22.

Jordanova V. K., A. Boonsiriseth, R. M. Thorne, and Y. Dotan (2003), Ring current asymmetry from global simulations using a high‐resolution electric field model, *J. Geophys. Res.*, *108*(A12), 1443, doi:10.1029/2003JA009993.

Jorgensen, A., M. Henderson, E. Roelof, G. Reeves, and H. Spence (2001), Charge exchange contribution to the decay of the ring current, measured by energetic neutral atoms (ENAs), *J. Geophys. Res.*, *106*(A2), 1931-1937.

Kalegaev, V. V., K. Yu Bakhmina, I. I. Alexeev, E. S. Belenkaya, Ya I. Feldstein, N. V. Ganushkina (2008), Ring Current Asymmetry during a Magnetic Storm, *Geomagnetism and Aeronomy*, *48*(6), 747–758.

Khazanov G. V., K. V. Gamayunov, V. K. Jordanova, and E. N. Krivorutsky (2002), A self-consistent model of interacting ring current ions and electromagnetic ion cyclotron waves, initial results: Waves and precipitating fluxes, *J. Geophys. Res.*, *107*(A6), 1085, doi:10.1029/2001JA000180.

Khazanov G. V., K. V. Gamayunov, and V. K. Jordanova (2003), Self‐consistent model of magnetospheric ring current and electromagnetic ion cyclotron waves: The 2–7 May 1998 storm, *J. Geophys. Res.*, *108*(A12), 1419, doi:10.1029/2003JA009856.

Kozyra, J. U., V. K. Jordanova, R. B. Horne, R. M. Thorne, (1997), Modeling of the Contribution of Electromagnetic Ion Cyclotron (EMIC) Waves to Stormtime Ring Current Erosion, in *Magnetic Storms, Geophys. Monogr. Ser, 98*. Edited by Bruce T. Tsurutani, et al., p.187, AGU, Washington D.C.

Kozyra, J., M.-C. Fok, E. Sanchez, D. Evans, D. Hamilton, and A. Nagy (1998), The role of precipitation losses in producing the rapid early recovery phase of the Great Magnetic Storm of February 1986, *J. Geophys. Res.*, *103*(A4), 6801-6814.

Kozyra, J. U., M. W. Liemohn, C. R. Clauer, A. J. Ridley, M. F. Thomsen, J. E. Borovsky, J. L. Roeder, V. K. Jordanova, and W. D. Gonzalez (2002), Multistep Dst development and ring current composition changes during the 4–6 June 1991 magnetic storm, *J. Geophys. Res.*, *107*(A8), 1224, doi:10.1029/2001JA000023.

Kozyra, J. U, and M. W. Liemohn (2003), Ring current energy input and decay, *Space Sci. Rev.*, *109*, 105–131.

Liemohn, M. W., J. U. Kozyra, V. K. Jordanova, G. V. Khazanov, M. F. Thomsen, and T. E. Cayton (1999), Analysis of early phase ring current recovery mechanisms during geomagnetic storms, *Geophys. Res. Lett.*, *26*(18) 2845–2848.

Liemohn, M. W., J. U. Kozyra, M. F. Thomsen, J. L. Roeder, G. Lu, J. E. Borovsky, and T. E. Cayton (2001), Dominant role of the asymmetric ring current in producing the stormtime Dst, *J. Geophys. Res., 106*(6), 10883–10904.

Liemohn, M. W., and J. U. Kozyra (2005), Testing the Hypothesis that charge exchange can cause a two-phase decay, in *Inner Magnetosphere Interactions: New Perspectives from Imaging, Geophys. Monogr. Ser., 159*, edited by J. L. Burch. M. Schulz, and H. Spence. pp. 167– 178, AGU, Washington, D. C.

Mitchell, D., K. Hsieh, C. Curtis, D. Hamilton, H. Voss, E. Roelof, and P. C:son-Brandt (2001), Imaging Two Geomagnetic Storms in Energetic Neutral Atoms, *Geophys. Res. Lett., 28*(6), 1151-1154.

Monreal MacMahon, R., and C. Llop (2008), Ring Current Decay Time Model during Geomagnetic Storms: a Simple Analitycal Approach, *Ann. Geophys.*, 26, 2543-2550.

O'Brien, T. P., and R. L. McPherron, and M. W. Liemohn (2002), Continued convection and the initial recovery of Dst, *Geophys. Res. Lett., 29*(23), 2143, doi: 10.1029/2002GL015556.

O'Brien, T. P., and R. L. McPherron (2000), An Empirical Phase Space Analysis of Ring Current Dynamics: Solar wind control of injection and decay, *J. Geophys. Res., 105*(A4), 7707–7719.

Ohtani, S., M. Nose, G. Rostoker, H. Singer, A. T. Y. Lui, and M. Nakamura (2001), Storm-substorm relationship: Contribution of the tail current to Dst, *J. Geophys. Res., 106*(A10), 21199–21209.

Pollock, C.J. et al. (2001), First Medium Energy Neutral Atom (MENA) images of Earth's magnetosphere during substorms and storm-time, *Geophys. Res. Lett., 28*(6), 1147-1150.

Pop, F. A., and K. H. Li (1993), Hyperbolic Relaxation as a Sufficient Condition of a Fully Coherent Ergodic Field, *Int. J. Theor. Phys., 32* (9), 1573-1583.

Prigancová, A., and Ya. I. Feldstein (1992), Magnetospheric Storm Dynamics in Terms of Energy Output Rate, *Planet. Space Sci.*, *40* (4), 581–588.

Sckopke, N. (1966), A General Relation between the Energy of Trapped Particles and the Disturbance Field near the Earth, *J. Geophys. Res.*, *71*, 3125–3130.

Sugiura, M., and T. Kamei (1991), Equatorial Dst index 1957–1986, ISGI, Saint-Maur-des-Fosses, France.

Søraas, F., K. Aarsnes, K. Oksavik, D.S. Evans (2002), Ring current intensity stimated from low-altitude proton observations, *J. Geosphys. Res*., *107*(A7), SMP 30-1, 1149, doi: 10.1029/2001JA000123

Søraas, F., K. Oksavik, K. Aarsnes, D. S. Evans, M. S. Greer (2003), Storm time equatorial belt - an ``image'' of RC behavior, *Geophys. Res. Lett*., 30(2), 24-1, 1052, doi: 10.1029/2002GL015636

Takahashi, S., T. Iyemori, and M. Takeda (1990), A Simulation of Storm-time Ring Current, *Planet. Space Sci.*, *38*(9), 1133–1141.

Tinsley, B. A., and S.-I. Akasofu (1982), A note on the lifetime of the ring current particles, *Planet. Space Sci*., 30, 733-740.

Walt, M., and H. Voss (2001), Losses of Ring Current Ions by Strong Pitch Angle Scattering, *Geophys. Res. Lett.*, *28*(20), 3839-3841.

Wang, C. B., J. K. Chao, and C.-H. Lin (2003), Influence of the Solar Wind Dynamic Pressure on the Decay and Injection of the Ring Current, *J. Geophys. Res.*, *108*(A9), 1341, doi:10.1029/2003JA009851.

Weygand J. M., R. L. McPherron (2006), Dependence of ring current asymmetry on storm phase, *J. Geophys. Res*., 111, A11221, doi:10.1029/2006JA011808.

# Figure Captions

**Figure 1.** Superposed epoch plot corresponding to the mean recovery phases of different subsets: from -100 to -150 nT (filled dots), -150 to -200 nT (empty dots), -200 to -250 nT (filled squares), -250 to -300 nT (empty squares) and less than -300 nT (crosses).

**Figure 2.** Exponential (blue dashed line) and hyperbolic decay (red solid line) fitting with the mean recovery phase (dots) of the storm subsets. The corresponding subset is indicated in each panel.

**Figure 3.** Characteristic recovery time from hyperbolic fitting versus the fitting parameter related to the intensity of the storm, $Dst_0$ (dots) and linear regression curve (solid line). The dashed line shows the linear regression keeping the point (-218.3 nT, 12.37 hours) out of the sample. The plus symbol corresponds to a new analysis of the subset #3 modifying the criterion for a negligible injection of energy to dips which do not exceed 5 per cent of $Dst_{peak}$ value (see text for details).

**Figure 1**

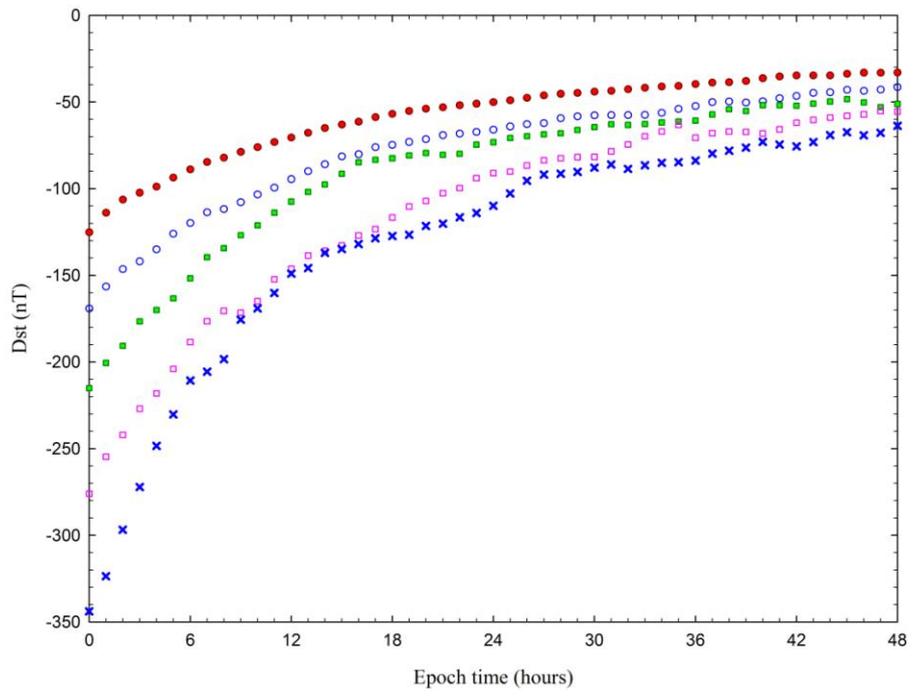

**Figure 2**

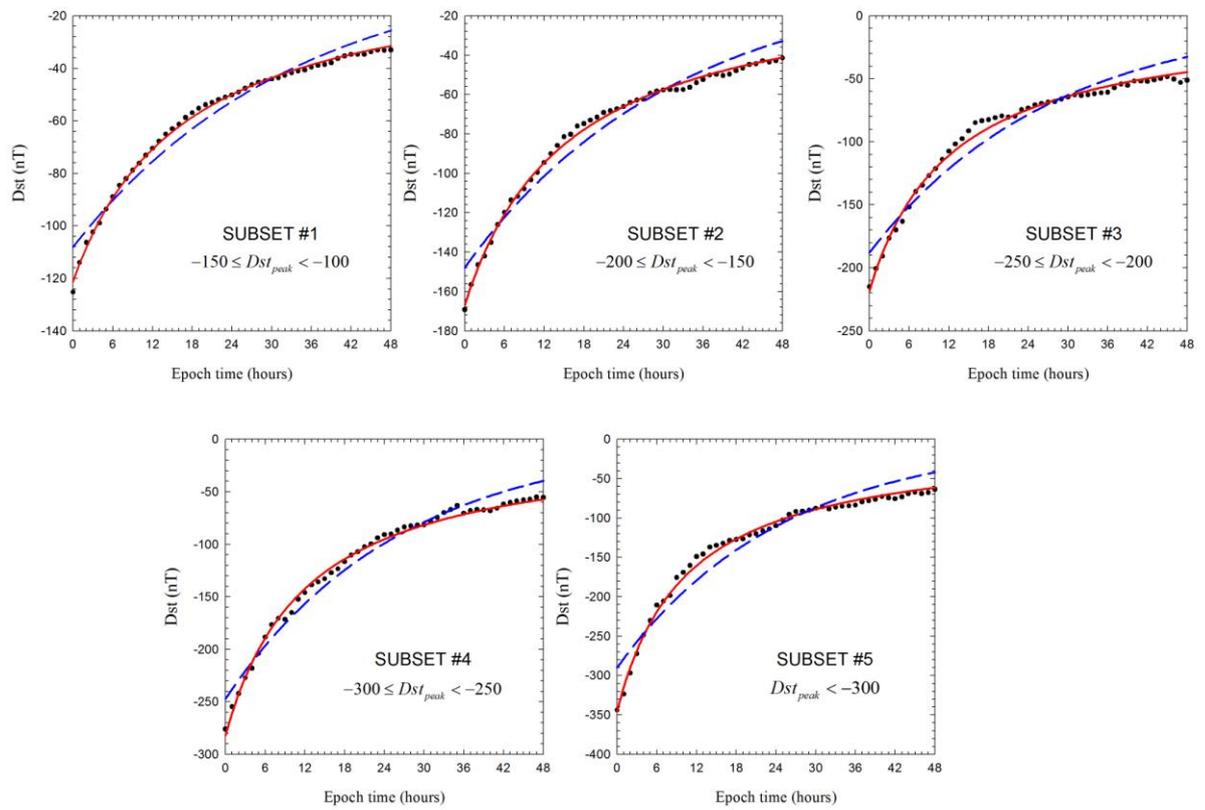

**Figure 3**

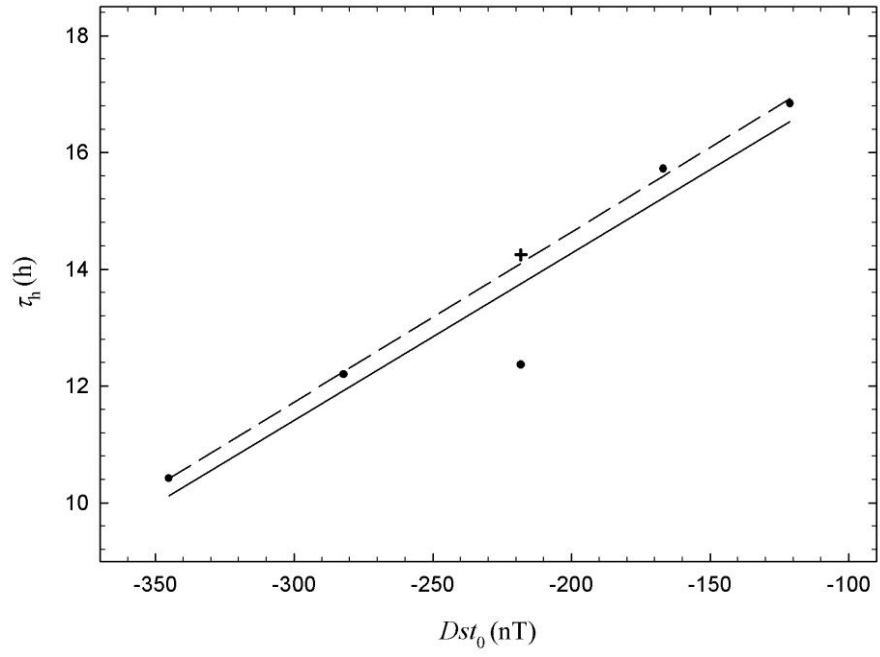